\documentstyle[pra,eqsecnum,aps]{revtex}

\begin{document}
\draft

\title{Problems and Aspects of Energy-Driven Wavefunction Collapse Models}
\author{Philip Pearle}
\address{Department of Physics, Hamilton College, Clinton, NY  13323}
\date{\today}

\maketitle

\begin{abstract}
{Four problematic circumstances are considered, involving models 
which describe dynamical wavefunction collapse toward energy 
eigenstates, for which it is shown that wavefunction collapse of macroscopic objects does not work properly.  
In one case, a common particle 
position measuring situation, the apparatus evolves to a 
superposition of macroscopically distinguishable states (does not collapse to one of them as it should) 
because each such particle/apparatus/environment state
has precisely the same energy spectrum. Second, assuming an experiment takes place involving collapse to one
of two possible outcomes which is permanently recorded, it is shown in general that 
this can only happen in the unlikely case that the two apparatus states corresponding to the two 
outcomes have disjoint energy spectra. Next, the progressive narrowing of the energy spectrum 
due to the collapse mechanism is considered. This has the effect of broadening the 
time evolution of objects as the universe evolves. Two examples, one involving a precessing spin, the 
other involving creation of an excited state followed by its decay, are presented in 
the form of paradoxes.  In both examples, 
the microscopic behavior predicted by standard quantum theory is significantly altered under energy-driven collapse, but    
this alteration is not observed by an apparatus when it is included in the 
quantum description. The resolution involves recognition 
that the statevector describing the apparatus does not collapse, but evolves to a superposition of  
macroscopically different states.}  
\end{abstract}

\pacs{03.65 Bz}

\section{Introduction}\label{Section I}

	Wavefunction collapse models alter Schr\"odinger's equation by adding 
a term which depends upon a stochastically fluctuating quantity.  The equation 
then drives a superposition of quantum states toward one or another state in the superposition (which 
state is realized depends upon the realization of the fluctuating quantity). Moreover, when all possible 
fluctuations and their probabilities are considered, the realized state appears as an outcome with 
the Born probability \cite{Pearle76,Pearle79}.

	Following seminal ideas involving collapse of particles to localized positions 
in the model of Ghirardi, Rimini and Weber \cite{GRW}, it became possible to 
construct a model, the Continuous Spontaneous Localization model (CSL)\cite{PearleCSL,GPR} 
which describes collapse based upon local density: particle number density, mass density 
\cite{PearleSquires,Exptl,GhirWeber} or (relativistic) energy density. This model entails e.g., rapid collapse 
of a superposition of macroscopically distinguishable 
spatially localized states (such as occurs in a measurement situation) 
to one of them. It does this because each state differs from the others in the superposition 
in its spatial distribution of energy density.  The CSL model, different as it is 
from standard quantum theory, has experimentally testable 
consequences\cite{Exptl,Penrose,Randomwalk}; so far it is consistent with experiment. 

	A long time ago, Bedford and Wang\cite{southafricans} proposed that that 
collapse based upon energy (not energy density as described above) could be 
viable.  More recently, Percival\cite{Percival} has constructed a stochastic 
energy-driven collapse model for microscopic systems but has not 
extended it to macroscopic systems so one cannot say how, e.g., 
the behavior of an apparatus is described in his model.  Most recently, 
Hughston\cite{Hughston} has given an elegant argument propounding 
an energy-driven collapse model, which has been followed by a number of papers 
\cite{AdlerHorwitz,others,HughstonBrody,Adler} exploring mathematical and physical consequences 
of this proposal. 

	Since the purpose of collapse models is to allow (the statevector of the modified) quantum theory to 
describe the localized world we see around us, it is necessary to have a mechanism 
whereby energy-driven collapse results in localized states of e.g., 
a macroscopic apparatus.  Hughston\cite{Hughston} has suggested that  
exchange of environment particles (air) with the apparatus by accretion/evaporation might achieve this result, 
and this has been explored by Adler\cite{Adler}. 

	However, after presenting the necessary energy-driven collapse formalism in Section II, we argue 
that it cannot (by this or any other mechanism) 
lead to spatially localized states in commonly occurring cases. Section III   
shows that this is so in a measurement situation, the 
position measurement of a superposition of two mutually translated particle states, when the particle, 
apparatus and environment are collectively described by the statevector. The reason is that 
the evolving superposed macroscopically distinguishable states have precisely the 
same energy spectra so one state is not singled out by energy-driven collapse, which is only sensitive to 
energy spectra differences. It is not sensitive to what 
does distinguish the different macroscopic states: 
their (generally degenerate) energy states have different phase factors, whose cancellation 
or augmentation, when projected into the position representation, gives rise to 
the spatial distinctiveness.

	In Section IV, after laying out the time-translation invariant properties 
of the formalism not immediately evident in Section II, the results are used to prove that,  
for an experiment which produces a permanent record, the associated apparatus states must have 
nonoverlapping energy spectra (which is most unlikely).   

	An aspect of energy-driven collapse is that it does not allow objects to change too rapidly. The reason is that 
each object has a smallest time $\tau$ which characterizes the time evolution of 
its fastest elements. Unless the state vector describing it contains a superposition of 
energy eigenstates with spread $\Delta E \geq \hbar/\tau$, it cannot evolve over $\tau$. 
However, in the energy-driven collapse model, {\it every} statevector has its energy bandwidth 
progressively narrowed as time wears on, to $\Delta E\approx (\lambda t)^{-1/2}$.   
Here,  $\lambda$ is 
a parameter in the model of dimension energy$^{-2}$time$^{-1}$, which characterizes 
the collapse rate and $t$ is the time that has elapsed since 
the energy-driven collapse began, presumably the beginning of the universe.    
Section V shows that any expectation value  
in standard quantum theory with $\tau<{\cal T}\equiv \hbar (\lambda t)^{1/2}$, 
has its time behavior ``smeared" so that, under energy-driven collapse,  
it evolves over the time interval ${\cal T}$.  We show that this is not only true for 
the ensemble of universes, but also true for a temporarily noninteracting subsystem of a single universe.   

	In Section VI we show the effect this has on a precessing spin and on an excitation/decay of
a bound state. The time parameters of the models, described by standard quantum theory,  
are chosen $<<{\cal T}$. The result is that, under energy-driven collapse,  the spin 
does not precess and nonexponential excitation/decay takes place over ${\cal T}$. 
However, when the apparatus is included in the statevector, the altered behavior is not recorded by it.  We couch this as  
a ``paradox," a contradiction between the description of the microscopic system alone, 
and the description of its measurement.  The resolution is that 
the measurement description does not produce collapse as it should.

\section{Energy Collapse Description}\label{Section II}

	The most transparent formulation of the energy-driven collapse model is as follows\cite{PearleNaples}. 
The statevector is assumed to obey the nonunitary evolution

\begin{equation}\label{2.1} 
|\psi,t\rangle_{B}=e^{-iHt}e^{-{1\over 4\lambda t}[B(t)-2\lambda t H]^{2}}|\psi,0\rangle.  
\end{equation}

\noindent In Eq. (2.1), H is the energy operator and $\lambda$ is the aforementioned constant parameter 
which characterizes the rate of energy collapse.  $B(t)$ 
is the stochastic variable whose time dependence is that of Brownian motion (i.e., it is continuous 
but not differentiable).  At any time t, it takes on the value $(B, B+dB)$ with probability

\begin{equation}\label{2.2} 
{\cal P}(B)dB={dB\over \sqrt{2\pi\lambda t}}\thinspace_{B}\langle\psi,t|\psi,t\rangle_{B}, 
\end{equation}

\noindent i.e., statevectors of largest norm occur with greatest probability.  Since the 
statevector (2.1) is not normed to 1, the expectation value of an operator $A$ in this state is

\begin{equation}\label{2.3}
\langle A\rangle_{B}(t) \equiv \thinspace_{B}\langle\psi,t|A|\psi,t\rangle_{B}/\thinspace_{B}\langle\psi,t|\psi,t\rangle_{B}.
\end{equation}

\noindent and the ensemble expectation value of an operator $A$ follows from Eqs. (2.2), (2.3):

\begin{equation}\label{2.4}
\langle A\rangle(t) =\int dB{\cal P}(B)\langle A\rangle_{B}(t)=
{\int dB\over \sqrt{2\pi\lambda t}}\thinspace_{B}\langle\psi,t|A|\psi,t\rangle_{B}.
\end{equation}

\noindent  All the conclusions about energy driven collapse in this paper follow from these equations. 

	For example, from Eq. (2.1) expressed in the energy basis, we may write 
	
\begin{equation}\label{2.5} 
\langle E,j|\psi,t\rangle_{B}=e^{-iEt}e^{-\lambda t[E-(B(t)/2\lambda t)]^{2}}\langle E,j|\psi,0\rangle  
\end{equation}

\noindent ($j$ is the energy degeneracy index).  As $t\rightarrow\infty$, in Eq. (2.5),
the gaussian $\rightarrow (\pi/\lambda t)^{1/2}	\delta [E-(B(t)/2\lambda t)]$, showing that the 
statevector collapses to an energy eigenstate (the value of the eigenstate 
determined by the asymptotic value of $B(t)/2\lambda t$).

\section{Position Measurement}\label{Section III}

	In most of this section we establish the structure of an initial state 
describing a particle in a superposition of two locations, accompanied by a 
position measuring apparatus immersed in a gaseous environment.  The evolution of this state 
in standard quantum theory is given by Eq. (3.9) and, in the energy driven 
collapse model, by Eq. (3.10). Eq. (3.11) shows that proper collapse does not take place. 

	Consider first an isolated particle which, at $t=0$, is in a localized state (for example, 
a gaussian wavepacket with some mean position, mean momentum and width).  The statevector of that 
particle, translated by the distance vector ${\bf a}$ but identical in every other respect, differs 
only in that the momentum eigenstates (and therefore energy eigenstates) of 
which it is composed are multiplied by the phase factor $\exp -i{\bf k}\cdot{\bf a}$. Label these 
two localized particle states by the index $\alpha=1,2$.  Their expansion in energy eigenstates of the 
Hamilton $H_{\rm part}$ may be written 

 \begin{equation}\label{3.1} 
|\phi^{\alpha},0\rangle =\sum_{kj}a_{kj}e^{i\theta_{kj}^{\alpha}}|\epsilon_{k},j\rangle 
\end{equation}

\noindent where $a_{kj}$ (which does {\it not} depend upon $\alpha$) and $\theta_{kj}^{\alpha}$ 
are real. For simplicity of notation we have written (3.1) as a sum, although the 
possible energy values $\epsilon_{k}$ ($=k^{2}/2m$) and the degeneracy index $j$ 
(which is the direction ${\bf k}/k$) are really continuous. 

	We assume that an appropriate beam splitter can put the particle in an 
initial state which is an arbitrary superposition 
$|\psi,0\rangle=\beta_{\alpha}|\phi^{\alpha},0\rangle$ with 
$\sum_{\alpha=1}^{2}|\beta_{\alpha}|^{2}=1$.  

	Consider next a position measuring apparatus together with environmental particles 
all around, but without the particle to be detected.  The initial state here is 

 \begin{equation}\label{3.2} 
|A,0\rangle =\sum_{mn}c_{mn}|\epsilon_{m}^{A},n\rangle
\end{equation}

\noindent where $|\epsilon_{m}^{A},n\rangle$ are energy eigenstates of the 
apparatus/environment Hamiltonian $H_{a/e}$ with energy $\epsilon_{m}^{A}$,  
$n$ is the degeneracy label, and $c_{mn}$ is generally a complex number. 

	Now, when the particle to be detected is brought together with the 
apparatus/environment, the energy eigenstates of the complete 
Hamiltonian, $H=H_{\rm part}+H_{\rm a/e}+H_{\rm int}\equiv H_{0}+H_{\rm int}$ must be considered.  
Lest the environment knock the particle away from the detector, take $H_{\rm int}$ 
not to include an environment-particle interaction (alternatively, one could   
choose the initial state so the particle reaches the apparatus before it is hit by an environment particle, 
but that complicates proofs).  Thus, $H_{\rm int}=\sum_{iapp}V_{i}({\bf X}_{iapp}-{\bf X})$, where ${\bf X}$ 
is the particle's position operator, 
$V_{i}$ is the interaction potential between the particle and each relevant 
apparatus particle (i.e., particles in 
the detector part of the apparatus, such as the gas in a geiger tube) whose position 
operators are ${\bf X}_{iapp}$. Thus, in particular, if $|{\bf x}\rangle$   
is an eigenstate of ${\bf X}$, then

\begin{equation}\label{3.3} 
  H_{\rm int}|{\bf x}\rangle=0 \thinspace\thinspace\thinspace\thinspace\thinspace\thinspace
  \hbox{when {\bf x} lies outside the particle detector}. 
\end{equation}
 
	It is necessary to write the initial state of particle together with the apparatus/environment in terms of 
the energy eigenstates of $H$. Given an energy eigenstate of the noninteracting particle 
$|\epsilon_{k},j\rangle$ (i.e., a plane wave state) and an energy 
eigenstate of the apparatus/environment $|\epsilon_{m}^{A},n\rangle$, there 
corrresponds a unique energy eigenstate of $H$, $|E_{km},j,n>$, satisfying the incoming 
Lippmann-Schwinger equation with energy $E_{km}=\epsilon_{k}+\epsilon_{m}^{A}$, 
which may be written 

\begin{mathletters}\label{all3.4}
\begin{eqnarray}
|E_{km},j,n>=&&|\epsilon_{k},j\rangle|\epsilon_{m}^{A},n\rangle+{1\over E_{km}+i\epsilon-H_{0}} H_{\rm int}|E_{km},j,n>\\
=&&|\epsilon_{k},j\rangle|\epsilon_{m}^{A},n\rangle+{1\over E_{km}+i\epsilon-H} H_{\rm int}
|\epsilon_{k},j\rangle|\epsilon_{m}^{A},n\rangle\\
\equiv&&|\epsilon_{k},j\rangle|\epsilon_{m}^{A},n\rangle+|\psi_{\rm int},k,m,j,n>.
\end{eqnarray}
\end{mathletters}

\noindent In Eq. (3.4c), $|\psi_{\rm int},k,m,j,n>$  
describes the various outcomes of detecting the particle by the apparatus along with 
scattering, accretion, excitation etc. involving the apparatus and environment. One may think of this 
as a scattering situation, with the apparatus a bound state of its constituent particles,  
and even the environment particles can be imagined as confined in a large box surrounding 
the apparatus so they too are part of the bound state. The analysis may then be considered 
under the rubric of multi-channel scattering\cite{NewtonTaylor}: the in-states 
$|\epsilon_{k},j\rangle|\epsilon_{m}^{A},n\rangle$ are then a complete set 
in their subspace of the Hilbert space.    

	The labels $k,m,j,n$ of 
$|E_{km},j,n>$ do not describe eigenvalues of operators which commute with $H$; they 
describe eigenvalues of operators which commute with $H_{\rm part}$ and $H_{\rm a/e}$.  
Nevertheless, it is well known from scattering theory that,  
as a consequence of (3.4), 

\begin{equation}\label{3.5} 
   \langle E_{k'm'},j',n'|E_{km},j,n\rangle=
  \langle\epsilon_{k'},j'|\langle\epsilon_{m'}^{A},n'|\epsilon_{k},j\rangle|\epsilon_{m}^{A},n\rangle
   =\delta_{k'k}\delta_{j'j}\delta_{m'm}\delta_{n'n}.
\end{equation} 

\noindent Eq. (3.5) can be derived algebraically by taking the scalar product of Eq. (3.4a) with 
(the primed version of) itself and   
utilizing the Low Equation\cite{Low1} (which also follows from (3.4a)) to eliminate 
all terms involving $|\psi_{\rm int},k,m,j,n>$.  It is most 
easily derived (utilizing $\exp{it[H_{0}-E_{km}]}|\epsilon_{k},j\rangle|\epsilon_{m}^{A},n\rangle=
|\epsilon_{k},j\rangle|\epsilon_{m}^{A},n\rangle$) by writing Eq. (3.4b) as 

\begin{mathletters}\label{3.6}
\begin{eqnarray}
|E_{km},j,n>=&&[1-i\int_{0}^{\infty}dt e^{-it[H-E_{km}-i\epsilon]}
H_{\rm int}e^{it[H_{0}-E_{km}]}]|\epsilon_{k},j\rangle|\epsilon_{m}^{A},n\rangle\\ 
=&&[1-i\int_{0}^{1/\epsilon} dt e^{-itH}H_{\rm int}e^{itH_{0}}]|\epsilon_{k},j\rangle|\epsilon_{m}^{A},n\rangle=
[1+\int_{0}^{1/\epsilon}d (e^{-itH}e^{itH_{0}})]|\epsilon_{k},j\rangle|\epsilon_{m}^{A},n\rangle\\ 
=&&\lim_{t\rightarrow\infty}e^{-itH}e^{itH_{0}}|\epsilon_{k},j\rangle|\epsilon_{m}^{A},n\rangle
\equiv \Omega^{+}|\epsilon_{k},j\rangle|\epsilon_{m}^{A},n\rangle. 
\end{eqnarray} 
\end{mathletters} 

\noindent Here, $\Omega^{+}$ is the M\slash \negthinspace\negthinspace \negthinspace oller matrix.   
 It is well known to be isometric, $\Omega^{+\dag}\Omega^{+}=1$ on the space of 
in-states, the proof of which we 
 now sketch for completeness. As shown above, so it is also readily shown from Eq. (3.4a) that 

\begin{equation}\label{3.7}
|\epsilon_{k},j\rangle|\epsilon_{m}^{A},n\rangle 
=\lim_{t\rightarrow\infty}e^{-itH_{0}}e^{itH}|E_{km},j,n>
=\Omega^{+\dag}|E_{km},j,n>,
\end{equation}

\noindent so $\Omega^{+\dag}\Omega^{+}|\epsilon_{k},j\rangle|\epsilon_{m}^{A},n\rangle
=|\epsilon_{k},j\rangle|\epsilon_{m}^{A},n\rangle$ follows from Eqs. (3.6), (3.7).  
Eq. (3.5) is an immediate consequence of the scalar product of (3.6c) with (the primed version of) itself, 
and use of $\Omega^{+\dag}\Omega^{+}=1$. 
        
	The $\alpha$th initial state can be written as a sum over eigenstates of $H$ as follows.  
From Eqs. (3.1), (3.2) and again using 
the form (3.6a), one sees that 
  
\begin{mathletters}
\label{all3.8}
\begin{eqnarray}
\sum_{kjmn}a_{kj}e^{i\theta_{kj}^{\alpha}}c_{mn}|E_{km},j,n>=&&|\phi^{\alpha},0\rangle|A,0\rangle
-i\int_{0}^{\infty}dt e^{-it[H-i\epsilon]}H_{\rm int}
\sum_{kj}a_{kj}e^{i\theta_{kj}^{\alpha}}e^{it\epsilon_{k}}|\epsilon_{k},j\rangle
\sum_{mn}c_{mn}e^{it\epsilon_{m}^{A}}|\epsilon_{m}^{A},n\rangle\\
=&&|\phi^{\alpha},0\rangle|A,0\rangle
-i\int_{0}^{\infty}dt e^{-it[H-i\epsilon]}H_{\rm int}
|\phi^{\alpha},-t\rangle |A, -t\rangle\\
=&&|\phi^{\alpha},0\rangle|A,0\rangle.  
\end{eqnarray}
\end{mathletters}

\noindent The second term in (3.8b) vanishes because $|\phi^{\alpha},0\rangle$ is a wavepacket outside the 
apparatus with mean momentum heading toward the apparatus.  Therefore,  $|\phi^{\alpha},-t\rangle$ $(t>0)$ is the 
wavepacket even further away from the apparatus, so (3.8c) follows from (3.3). (This 
result is well known for potential scattering\cite{Low2}.)

	Eqs. (3.5) and (3.8c) are what we need as we now consider the time evolution. 
	 
	In standard quantum theory, using Eq. (3.8c), the time evolution of the $\alpha$th initial state is  
	
\begin{equation}\label{3.9} 
|\phi^{\alpha},0\rangle|A,0\rangle\rightarrow|\Psi^{\alpha},t\rangle=
\sum_{kjmn}a_{kj}e^{i[\theta_{kj}^{\alpha}-E_{km}t]}c_{mn}|E_{km},j,n\rangle
\end{equation}
	  
\noindent where $|\Psi^{1},t\rangle$,  $|\Psi^{2},t\rangle$ eventually describe the apparatus as having detected 
the particle's location (and also describe the battering by the environment particles of the 
apparatus and each other) and having indicated this by a macroscopically distinguishable spatially localized feature.  

	According to the energy-driven collapse model, Eq. (2.1), and once again employing Eq. (3.8c), we see that the  
initial superposed state $|\Psi,0\rangle=\sum_{\alpha =1}^{2}\beta_{\alpha}|\phi^{\alpha},0\rangle|A,0\rangle$ 
evolves into the (unnormalized) state 

\begin{mathletters}
\label{all3.10}	 
\begin{eqnarray} 
|\Psi,t\rangle_{B}=&&\sum_{\alpha =1}^{2}\beta_{\alpha}
\sum_{kjmn}e^{i[\theta_{kj}^{\alpha}-E_{km}t]} e^{-{1\over 4\lambda t}[B(t)-2\lambda t E_{km}]^{2}}
a_{kj}c_{mn}|E_{km},j,n\rangle\\ 
\equiv&& \sum_{\alpha =1}^{2}\beta_{\alpha}|\Psi^{\alpha},t\rangle_{B}
 \end{eqnarray}
\end{mathletters} 

\noindent for some $B(t)$. (For simplicity we have taken $t=0$  
as the time the collapse process begins: replacement of $t$ by $t+T$ in the gaussian in Eq. (3.10a) 
does not affect the result (3.11).)   
Assuming that the energy-driven collapse does not destroy the functioning of 
the apparatus, the orthogonal states $|\Psi^{1},t\rangle_{B}$, $|\Psi^{2},t\rangle_{B}$
each describe the apparatus as possessing a spatially localized macroscopically distinguishable feature.  

	However, the squared amplitudes of the states $\beta_{\alpha}|\Psi^{\alpha},t\rangle_{B}$ in the superposition (3.10) are 
(apart from an overall normalization factor which does not affect their ratio):
	 
\begin{equation}\label{3.11}
|\beta_{\alpha}|^{2}\thinspace_{B}\langle\Psi^{\alpha},t|\Psi^{\alpha},t\rangle_{B}=
|\beta_{\alpha}|^{2}\sum_{kjmn}a_{kj}^{2}|c_{mn}|^{2}e^{-{1\over 2\lambda t}[B(t)-2\lambda t E_{km}]^{2}} 
\end{equation}

\noindent on account of the orthonormality (3.5). But, the $\alpha$ dependence is 
absent from the sum in Eq. (3.11).  Thus, there is no collapse 
to apparatus localized states since the superposed states at time $t$ are in the same proportion, 
 $|\beta_{2}|^{2}/|\beta_{1}|^{2}$, as they were initially. 
 
 \section{Energy-Driven Collapse and Permanent Records}\label{IV}
 
 	According to Eq. (2.1), at time $t$, each value of $B(t)$ corresponds to a different possible universe in 
 the ensemble of universes, all of which evolved from a single statevector $|\psi,0\rangle$ 
 starting at time $t=0$. If collapse takes place as purported, 
 each $B(t)$ should characterize a recognizable state of the universe, in the sense that each 
 macroscopic object in a universe is not in a superposed state.  That a single number $-\infty<B(t)<\infty$ 
 can characterize the myriad possible recognizable universes which evolve from an initial 
 state $|\psi,0>$ seems unlikely, and raises the suspicion that collapse cannot take place as purported. 
 
 	In subsection A, time-translation properties are displayed, 
showing how the statevector and probability characterized by $B(t)$ can be expressed 
in terms of $B(t_{0})$ at time $t_{0}$, and not just in terms of $B(0)=0$ at time 0, 
as given in Eqs. (2.1), (2.2).  

	In subsection B, any experiment is considered whose 
outcome, one of two possible values, is macroscopically permanently recorded at time $t_{0}$ (i.e., 
the record is unaltered for $t>t_{0}$).  It is shown that 
this may only be possible if there is no overlap in the energy spectra at time $t_{0}$ corresponding to 
different states describing different macroscopic outcomes.  Since there is no reason 
why a measurement should lead to such a bifurcation of the energy spectrum, it must be 
concluded that there cannot be collapse to a state which describes a unique macroscopic 
permanent record.  
 
\subsection{Time Translation Invariant Description} 

	Corresponding to Eq. (2.1), it is useful to define 
the (unnormalized) statevector at time $t$ 

\begin{equation}\label{4.1} 
|\psi,t\rangle_{B(t), B(t_{0})}\equiv e^{-iH(t-t_{0})}e^{-{1\over 4\lambda(t-t_{0})}[B(t)-B(t_{0})-2\lambda (t-t_{0}) H]^{2}}
|\psi,t_{0}\rangle_{B(t_{0})}
\end{equation}

\noindent which evolves from the (unnormalized) statevector $|\psi,t_{0}\rangle_{B(t_{0})}$ at time $t_{0}$   
(we take $B(0)=0$ and $|\psi,t_{0}\rangle_{B(t_{0}),0}\equiv|\psi,t_{0}\rangle_{B(t_{0})}$ is 
given by Eq. (2.1)).  
 
 	Corresponding to Eq. (2.2), given $B(t_{0})$ (and therefore also the associated  
statevector $|\psi,t_{0}\rangle_{B(t_{0})}$), 
the conditional probability that, at time $t$, $B$ lies in the interval 
$(B(t), B(t)+dB(t))$ is:

\begin{equation}\label{4.2} 
{\cal P}(B(t),t|B(t_{0}),t_{0})dB(t)={dB(t)\over \sqrt{2\pi\lambda (t-t_{0})}}
{\thinspace_{B(t),B(t_{0})}\langle\psi,t|\psi,t\rangle_{B(t),B(t_{0})}\over 
\thinspace_{B(t_{0})}\langle\psi,t_{0}|\psi,t_{0}\rangle_{B(t_{0})}}. 
\end{equation}

	It is necessary to show that Eqs. (4.1), (4.2) are consistent with Eqs. (2.1), (2.2).
	
	Consider first the normalized statevector, which we shall denote by a prime.  
From Eq. (4.1),

\begin{equation}\label{4.3} 
|\psi,t\rangle'_{B(t), B(t_{0})}\equiv{|\psi,t\rangle_{B(t), B(t_{0})}\over 
\thinspace_{B(t), B(t_{0})}\langle\psi,t|\psi,t\rangle_{B(t), B(t_{0})}^{1/2}}=
 {e^{-iH(t-t_{0})}e^{[(B(t)-B(t_{0}))H-\lambda (t-t_{0}) H^{2}]}|\psi,t_{0}\rangle_{B(t_{0})}\over
 \thinspace_{B(t_{0})}\langle\psi,t_{0}|e^{2[(B(t)-B(t_{0}))H-\lambda (t-t_{0}) H^{2}]}|\psi,t_{0}\rangle_{B(t_{0})}^{1/2}}.
\end{equation}

\noindent (In Eq. (4.3), the term  
$\exp{-[4\lambda(t-t_{0})]^{-1}[B(t)-B(t_{0})]^{2}}$, arising from the squared bracket in (4.1)'s exponent, 
factors out of numerator and denominator and thus cancels, 
while the remaining terms from the squared bracket have 
a common factor $4\lambda(t-t_{0})$ which cancels with the premultiplying 
inverse of this factor).  

	The numerator of Eq. (4.3) has a purely linear dependence upon 
$B(t)-B(t_{0})$ and upon $(t-t_{0})$.  
Because of this, writing out $|\psi,t_{0}\rangle_{B(t_{0})}$ in (4.3) in terms of 
$|\psi,0\rangle$ using (2.1), cancelling the exponential factors 
$\exp{-[4\lambda t_{0}]^{-1}[B(t_{0})]^{2}}$ which then appear in numerator and denominator, 
and combining the remaining exponents

\[ \{-iH(t-t_{0})+(B(t)-B(t_{0}))H-\lambda (t-t_{0}) H^{2}\}+ 
\{-iHt_{0}+B(t_{0})H-\lambda t_{0} H^{2}\}=-iHt+B(t)H-\lambda t H^{2},
\]

\noindent results in 

\begin{equation}\label{4.4} 
|\psi,t\rangle'_{B(t), B(t_{0})}=
 {e^{-iHt} e^{B(t)H-\lambda t H^{2}}|\psi,0\rangle\over
 \langle\psi,0|e^{2[B(t)H-\lambda t H^{2}]}|\psi,0\rangle}
 ={e^{-iHt} e^{-{1\over 4\lambda t}[B(t)-2\lambda t H]^{2}}|\psi,0\rangle\over
 \langle\psi,0|e^{-{1\over 2\lambda t}[B(t)-2\lambda t H]^{2}}|\psi,0\rangle}.
\end{equation}

\noindent We see in Eq. (4.4) that the dependence upon $B(t_{0})$ and $t_{0}$ has disappeared and that the 
normalized statevector (4.3), constructed from (4.1), is 
the same normalized statevector that is constructed from (2.1). 

	Consider next the probability that $B$ takes on the value 
$B(t)$ at time $t$, given that it has the value 0 at time 0,  From Eqs. (4.2), (4.1) 
and the usual rule for compounding probabilities:

\begin{eqnarray}\label{4.5} 
{\cal P}(B(t),t|0,0)dB(t)&&=dB(t)\int {\cal P}(B(t),t|B(t_{0}),t_{0})dB(t_{0}){\cal P}(B(t_{0}),t_{0}|0,0)\nonumber\\ 
&&={dB(t)\over \sqrt{2\pi\lambda (t-t_{0})}}\int 
{\thinspace_{B(t),B(t_{0})}\langle\psi,t|\psi,t\rangle_{B(t),B(t_{0})}\over 
\thinspace_{B(t_{0})}\langle\psi,t_{0}|\psi,t_{0}\rangle_{B(t_{0})}}
{dB(t_{0})\over \sqrt{2\pi\lambda t_{0}}}\thinspace_{B(t_{0})}\langle\psi,t_{0}|\psi,t_{0}\rangle_{B(t_{0})}\nonumber\\
&&={dB(t)\over \sqrt{2\pi\lambda (t-t_{0})}}\int {dB(t_{0})\over \sqrt{2\pi\lambda t_{0}}}
\thinspace_{B(t_{0})}\langle\psi,t_{0}|
e^{-{1\over 2\lambda(t-t_{0})}[B(t)-B(t_{0})-2\lambda (t-t_{0}) H]^{2}}|\psi,t_{0}\rangle_{B(t_{0})}\nonumber\\
&&= {dB(t)\over \sqrt{2\pi\lambda (t-t_{0})}}\int {dB(t_{0})\over \sqrt{2\pi\lambda t_{0}}}
\langle\psi,0|
e^{-{1\over 2\lambda (t-t_{0})}[B(t)-B(t_{0})-2\lambda (t-t_{0}) H]^{2}}
e^{-{1\over 2\lambda t_{0}}[B(t_{0})-2\lambda t_{0} H]^{2}}|\psi,0\rangle\nonumber\\
&&={dB(t)\over \sqrt{2\pi\lambda t}}\langle\psi,0|
e^{-{1\over 2\lambda t}[B(t)-2\lambda t H]^{2}}|\psi,0\rangle
 \end{eqnarray}

\noindent which is the same as Eq. (2.2).  

	This time translation invariance, especially evident in the stochastic differential Schr\"odinger 
equation (which we have not bothered to reproduce) whose solution is 
(4.1), seems to have misled workers into overlooking the 
important effects of the cumulative narrowing of the 
energy spectrum as the universe evolves (see sections V, VI).  Because of this,  one 
cannot take an initial statevector $|\psi,t_{0}\rangle_{B(t_{0})}$ to be just any statevector (as one is free to do in 
standard quantum theory or CSL) if $t_{0}=T$ is the age of the universe.  The evolution of 
the statevector from time 0 when the universe began to 
time $T$ restricts the spectrum of $|\psi,T\rangle_{B(T)}$, which must be respected if the theory is to be 
consistently applied. For example, it may not be consistent to consider as an initial condition 
that an electron is in an excited atomic state at time $T$. Although the electron is in an energy 
eigenstate (and therefore the energy spread is zero) of the atomic Hamiltonian, it is not in an energy 
eigenstate of the complete Hamiltonian (including the radiation field) and, for sufficiently 
large coupling (sufficiently short lifetime), the actual energy spread of this initial state may be larger than 
is allowed in the universe at time $T$.   

\subsection{Permanent Records} 

	Now, consider an experiment performed at a time earlier than $t_{0}$ 
which, for definiteness, has two equally likely outcomes, say 1 and -1.  Suppose further that the  
apparatus records the result at time $t_{0}$, which is permanent thereafter in all universes.  
A word on terminology: 
we are using ``universe" for what is described by the statevector $|\psi,t\rangle_{B(t)}$ corresponding to  
a particular $B(t)$, and ``universes" to characterize the set of these.  As far as the argument 
here is concerned, one may regard a universe as consisting of just the matter required for the experiment, or 
it may be so large as to represent the real universe (in this case, in order that all universes 
have the experiment going on as described, one might imagine that, somehow, at the beginning of the universe 
an isolated box containing the apparatus came into existence, with a clock 
set to start the experiment at a prescribed time).
 
	If collapse takes place as is supposed, one may partition the values of $B(t)$ for $t\geq t_{0}$
into sets $\Sigma_{1}(t)$ ($\Sigma_{1}(t)$ consists of 
the set $\{ B_{1}(t)\}$ for which $|\psi,t\rangle_{B_{1}(t)}$ describes the experimental outcome 1), 
$\Sigma_{-1}(t)$ (similarly defined) and $\Sigma_{0}(t)$ (covering the remaining values of 
$B(t)$ which do not lie in $\Sigma_{1}(t)$ or $\Sigma_{-1}(t)$ and which therefore must have 
negligible probability of occurring). Then, for example, if $B_{1}(t_{0})$ characterizes a universe 
(result 1 at time $t_{0}$), it is certain to evolve to a universe at time $t$ characterized by some $B_{1}(t)$ and 
it has 0 probability to evolve to a universe characterized by a $B_{-1}(t)$ or $B_{0}(t)$.
Thus, from the expression (4.2) for the conditional probability, we have:

\begin{mathletters}\label{4.6}	 
\begin{eqnarray}
&&\int_{\Sigma_{1}(t)}{dB(t)\over \sqrt{2\pi\lambda (t-t_{0})}}
{\thinspace_{B(t),B_{1}(t_{0})}\langle\psi,t|\psi,t\rangle_{B(t),B_{1}(t_{0})}\over 
\thinspace_{B_{1}(t_{0})}\langle\psi,t_{0}|\psi,t_{0}\rangle_{B_{1}(t_{0})}}\approx 1,
\int_{\Sigma_{1}(t)}{dB(t)\over \sqrt{2\pi\lambda (t-t_{0})}}
{\thinspace_{B(t),B_{-1}(t_{0})}\langle\psi,t|\psi,t\rangle_{B(t),B_{-1}(t_{0})}\over 
\thinspace_{B_{-1}(t_{0})}\langle\psi,t_{0}|\psi,t_{0}\rangle_{B_{-1}(t_{0})}}\approx 0\\
&&\int_{\Sigma_{-1}(t)}{dB(t)\over \sqrt{2\pi\lambda (t-t_{0})}}
{\thinspace_{B(t),B_{-1}(t_{0})}\langle\psi,t|\psi,t\rangle_{B(t),B_{-1}(t_{0})}\over 
\thinspace_{B_{-1}(t_{0})}\langle\psi,t_{0}|\psi,t_{0}\rangle_{B_{-1}(t_{0})}}\approx 1,
\int_{\Sigma_{-1}(t)}{dB(t)\over \sqrt{2\pi\lambda (t-t_{0})}}
{\thinspace_{B(t),B_{1}(t_{0})}\langle\psi,t|\psi,t\rangle_{B(t),B_{1}(t_{0})}\over 
\thinspace_{B_{1}(t_{0})}\langle\psi,t_{0}|\psi,t_{0}\rangle_{B_{1}(t_{0})}}\approx 0\\
&&\int_{\Sigma_{0}(t)}{dB(t)\over \sqrt{2\pi\lambda (t-t_{0})}}
{\thinspace_{B(t),B_{1}(t_{0})}\langle\psi,t|\psi,t\rangle_{B(t),B_{1}(t_{0})}\over 
\thinspace_{B_{1}(t_{0})}\langle\psi,t_{0}|\psi,t_{0}\rangle_{B_{1}(t_{0})}}\approx 0,
\int_{\Sigma_{0}(t)}{dB(t)\over \sqrt{2\pi\lambda (t-t_{0})}}
{\thinspace_{B(t),B_{-1}(t_{0})}\langle\psi,t|\psi,t\rangle_{B(t),B_{-1}(t_{0})}\over 
\thinspace_{B_{-1}(t_{0})}\langle\psi,t_{0}|\psi,t_{0}\rangle_{B_{-1}(t_{0})}}\approx 0
 \end{eqnarray}
\end{mathletters}

\noindent The ``$\approx$" is in these equations because a  
probability can differ from 0 or 1 by a negligible but nonzero amount and the theory still has  
acceptable behavior.  

	The scalar products of the statevectors in Eqs. (4.6) may be written in the form of Eq. (4.1) and 
numerators and denominators may be expressed in the energy basis:

\begin{mathletters}\label{4.7}	 
\begin{eqnarray}
{\thinspace_{B(t),B(t_{0})}\langle\psi,t|\psi,t\rangle_{B(t),B(t_{0})}\over 
\thinspace_{B(t_{0})}\langle\psi,t_{0}|\psi,t_{0}\rangle_{B(t_{0})}}&&=
{\sum_{j}\int dE |\langle E,j|\psi,t_{0}\rangle_{B(t_{0})}|^{2}
e^{-{1\over 2\lambda (t-t_{0})}[B(t)-B(t_{0})-2\lambda (t-t_{0}) E]^{2}}\over 
\sum_{j}\int dE |\langle E,j|\psi,t_{0}\rangle_{B(t_{0})}|^{2}}\\
&&\equiv\int dE e^{-{1\over 2\lambda (t-t_{0})}[B(t)-B(t_{0})-2\lambda (t-t_{0}) E]^{2}}\rho_{B(t_{0})}(E) 
\end{eqnarray}
\end{mathletters}

\noindent ($j$ is the degeneracy index for the energy eigenstates) where 

\[\rho_{B(t_{0})}(E)\equiv {\sum_{j}|\langle E,j|\psi,t_{0}\rangle_{B(t_{0})}|^{2}\over 
\sum_{j}\int dE |\langle E,j|\psi,t_{0}\rangle_{B(t_{0})}|^{2}}
\]
 
\noindent is the unity normalized energy spectrum of the universe at time $t_{0}$   
characterized by $B(t_{0})$.  

	We now show that Eqs. (4.6) can only be satisfied if $\rho_{B_{1}(t_{0})}(E)$, $\rho_{B_{-1}(t_{0})}(E)$ 
have disjoint support in $E$.  We apply Schwarz's inequality 

\[ \bigg[\int d{\bf x}f^{2}({\bf x})\int d{\bf x}g^{2}({\bf x})\bigg]^{1/2}\geq \int d{\bf x} f({\bf x})g({\bf x}) \]

\noindent (where $\int d{\bf x}f^{2}({\bf x})\approx 0$ and $\int d{\bf x}g^{2}({\bf x})\leq 1$ and $d{\bf x}=dB(t)dE$)  
to the product of the two equations in 
each of Eqs. (4.6a), (4.6b), (4.6c) expressed in the form (4.7b), obtaining:

\begin{equation}\label{4.8}
0\approx \int_{\Sigma_{i}}{dB(t)\over \sqrt{2\pi\lambda (t-t_{0})}}\int dE
e^{-{1\over 4\lambda (t-t_{0})}[B(t)-B_{1}(t_{0})-2\lambda (t-t_{0}) E]^{2}} 
e^{-{1\over 4\lambda (t-t_{0})}[B(t)-B_{-1}(t_{0})-2\lambda (t-t_{0}) E]^{2}}
\rho^{1/2}_{B_{1}(t_{0})}(E)\rho^{1/2}_{B_{-1}(t_{0})}(E)   
\end{equation}

\noindent where $\Sigma_{i}$ is any one of $\Sigma_{1}$, $\Sigma_{-1}$, $\Sigma_{0}$.  Adding the three 
expressions (4.8) together, the range of $B(t)$ becomes the whole line so the integral over 
$B(t)$ can then be performed, resulting in 

\begin{equation}\label{4.9}
0\approx e^{-{1\over 8\lambda (t-t_{0})}[B_{1}(t_{0})-B_{-1}(t_{0})]^{2}} 
\int dE \rho^{1/2}_{B_{1}(t_{0})}(E)\rho^{1/2}_{B_{-1}(t_{0})}(E).    
\end{equation}

	No matter how large is $[B_{1}(t_{0})-B_{-1}(t_{0})]^{2}$, for large enough 
$t$ the exponential in (4.9) is $\approx 1$.   Thus the integral must vanish and so, for each $E$, one (or both) of  
$\rho_{B_{1}(t_{0})}(E)$, $\rho_{B_{-1}(t_{0})}(E)$ must vanish (or be negligibly small). 

	However, there is no mechanism whereby a measurement ``splits" the energy spectrum 
of the universe (whether the universe is
just the apparatus or the real universe) into disjoint sets associated with each outcome. 
We conclude that the premise of this argument, that the energy-driven collapse mechanism 
will produce macroscopic states, each corresponding to a single recorded experimental outcome 
(and not produce a superposition of such states), is false.

\section{Secular Decrease in Energy Bandwidth and Consequences}\label{V}

	In what follows, for definiteness we shall consider that 
``universe" refers to the actual universe, and ``universes" describes 
an ensemble of which only one (that we inhabit) is actually realized. 
 
 	If one expands the initial statevector of the universe $|\psi,0\rangle$ in energy eigenstates, 
$|\psi,0\rangle =\int dE C_{j}(E)|E,j\rangle$ ($j$ is 
the degeneracy index), the statevector evolution (2.1) of a single universe 
characterized by $B(t)$ may be written using Eq. (2.5) as  
 
\begin{equation}\label{5.1} 
|\psi,t\rangle_{B}=\sum_{j}\int dE C_{j}(E)e^{-iEt}e^{-\lambda t[E-(B(t)/2\lambda t)]^{2}}|E,j\rangle. 
\end{equation}

\noindent Thus, for {\it each} statevector, the initial energy spectrum $\sum_{j}|C_{j}(E)|^{2}$ is 
increasingly narrowed as time wears on by being multiplied by a gaussian of energy spread $\sim (\lambda t)^{-1/2}$ (and 
mean value $B(t)/2\lambda t)$.  

	Consequently, rapid time evolution of anything in {\it any} universe is 
restricted. Supppose that the universe has evolved for 
time $T$, and we consider physical phenomena over a relatively short time interval 
thereafter, $(0, t-T)$, where $t-T<<T$.  Then the gaussian in Eq. (5.1) is not much different at time $t$ 
than at time $T$ over this time interval 
(i.e., ${\cal T}\equiv (\lambda t)^{1/2}\approx (\lambda T)^{1/2}$, and $B(t)/2{\cal T}^{2}\approx B(T)/2{\cal T}^{2}$), so Eq. (5.1) 
becomes approximately 

\begin{equation}\label{5.2}
|\psi,t\rangle_{B}\approx e^{-iH(t-T)}|\psi, T\rangle_{B}.  
\end{equation}
 
\noindent Since the (approximate) evolution (5.2) has the usual form 
and $|\psi,T\rangle_{B}$ has energy spread no greater than $\approx \hbar/{\cal T}$, 
according to the usual time-energy uncertainty relation we expect that the characteristic 
evolution time of any physical system in the so-restricted universe, 
e.g., any ``pulse" behavior, can be no shorter than ${\cal T}$. 

	In Section VA we consider the ensemble of universes. The above argument for the restriction on pulse 
behavior is made more precise, 
in the expression Eq. (5.4c) for the ensemble expectation value of any 
operator $V$. Also, from this expression, it is shown that the ensemble 
energy spectrum at any time $t$ is unchanged from that at time 0.  However,   
the energy spread restriction $\Delta E\leq \hbar/{\cal T}$ on each universe does manifest itself,  in 
the (approximate) vanishing  of the density matrix off-diagonal energy basis elements
when their difference exceeds $\approx\Delta E$. It is 
also shown that, were there a noninteracting subsystem present in each of the 
ensemble of universes (one may imagine that 
a physical system is present in $|\psi,0\rangle$ which somehow remains isolated 
for all time), that it alone also obeys Eq. (5.4c).  

	However, we only have access to one universe (one $B(t)$), not the ensemble.  
In Section VB it is argued that, in a {\it single} 
universe, when the rest of the universe is traced over,  
Eq. (5.4c) holds to a good approximation for a subsystem which may have interacted with the 
rest of the universe in the past but is presently not interacting with it.  
This allows us to discuss experiments in the single universe which are ``turned on' at time $T$, 
two examples of which are treated in Section VI. Since it is a single universe, if collapse takes place properly, 
the result of an experiment should entail a unique outcome, but we shall see that this is not the case.    
 	 
\subsection{Ensemble Behavior}

	Here we show that that the ensemble of universes has the same energy spectrum as $|\psi,0\rangle$.   
(Even though the energy spread of each universe is narrowed to $\leq \hbar/{\cal T}$, 
each universe's energy spread has a different mean value $\sim B(t)$, and these means are spread out 
over the whole real line.)   
However, the ``pulse" constraint discussed above, since it holds for all statevectors, holds for the ensemble.
  
	To see this formally, we write Eq. (2.1) 
in still another way, in terms of the statevector $|\psi,t\rangle$ which evolves from $|\psi,0\rangle$ 
under ordinary Schr\"odinger dynamics:

\begin{mathletters}
\begin{eqnarray}\label{5.3}
|\psi,t\rangle_{B}=&&e^{-{1\over 4\lambda t}[B(t)-2\lambda t H]^{2}}|\psi,t\rangle\\
=&&{1\over \sqrt{4\pi}}\int_{-\infty}^{\infty}d\eta e^{-\eta^{2}/4}e^{-iB\eta /\sqrt{4\lambda t}}
e^{i\sqrt{\lambda t}\eta H}|\psi,t\rangle\\
=&&{1\over \sqrt{4\pi}}\int_{-\infty}^{\infty}d\eta 
e^{-\eta^{2}/4}e^{-iB\eta /\sqrt{4\lambda t}}|\psi,t-\sqrt{\lambda t}\eta\rangle
\end{eqnarray}
\end{mathletters}

\noindent where the Fourier transform of the gaussian has been employed in Eq. (5.3b) to obtain an expression 
linear in $H$.  
Thus, according to Eq. (5.3c), $|\psi,t\rangle_{B}$ may be viewed as a "time-smearing" superposition of  
$|\psi,t\rangle$'s over a range of $t$ of order ${\cal T}$, with a gaussian weight 
and a $B$-dependent phase. 

	The ensemble average $\langle V\rangle$, of any 
physical quantity $V$, is found by   
putting Eq. (5.3c) into Eq. (2.4):

\begin{mathletters}
\begin{eqnarray}\label{5.4}
\langle V\rangle(t)=&&{1\over \sqrt{2\pi \lambda t}}\int_{-\infty}^{\infty}dB\thinspace_{B}\langle\psi,t|V|\psi,t\rangle_{B}\\
=&&{1\over4\pi}\int_{-\infty}^{\infty}d\eta \int_{-\infty}^{\infty}d\eta '
e^{-\eta^{2}/4}e^{-\eta '^{2}/4}\langle\psi,t-\sqrt{\lambda t}\eta '|V|\psi,t-\sqrt{\lambda t}\eta\rangle
 {1\over \sqrt{2\pi \lambda t}}\int_{-\infty}^{\infty}dB e^{-iB(\eta-\eta ')/\sqrt{4\lambda t}}\\
 =&&{1\over \sqrt{2\pi}}\int_{-\infty}^{\infty}d\eta e^{-\eta^{2}/2}
 \langle\psi,t-\sqrt{\lambda t}\eta |V|\psi,t-\sqrt{\lambda t}\eta\rangle.
\end{eqnarray}
\end{mathletters}

	Eq. (5.4c) makes precise the discussion in the introduction above. Any 
pulse-like behavior of $\langle V\rangle (t)$ in standard quantum theory is 
``smeared" over a time interval $\approx {\cal T}$ (with ${\cal T}\approx (\lambda T)^{1/2}$ for $|t-T|<<T$). 
	 
	If $V=F(H)$ is an arbitrary function of the energy, it follows from Eq. (5.4c) that  

\begin{equation}\label{5.5}
\langle F(H)\rangle(t)={1\over \sqrt{2\pi}}\int_{-\infty}^{\infty}d\eta e^{-\eta^{2}/2}
\langle\psi,0|e^{iH(t-\sqrt{\lambda t}\eta)}F(H)e^{-iH(t-\sqrt{\lambda t}\eta)}|\psi,0\rangle=\langle\psi,0|F(H)|\psi,0\rangle.
\end{equation}

\noindent i.e., the energy distribution of the ensemble of statevectors is always equal to its initial energy distribution.

	However, that is not all there is to say with regard to the energy. The 
density matrix in the energy basis may be written, considering Eq. (5.4c), as 

\begin{mathletters}
\begin{eqnarray}\label{5.6}
\langle E'|\rho|E\rangle=&&{1\over \sqrt{2\pi}}\int_{-\infty}^{\infty}d\eta e^{-\eta^{2}/2}
 \langle E'|\psi,t-\sqrt{\lambda t}\eta\rangle\langle\psi,t-\sqrt{\lambda t}\eta|E\rangle\\
 =&&{1\over \sqrt{2\pi}}\int_{-\infty}^{\infty}d\eta e^{-\eta^{2}/2}e^{i(E'-E)\sqrt{\lambda t}\eta} 
 \langle E'|\psi,t\rangle\langle\psi,t|E\rangle\\
 =&&e^{-\lambda t(E'-E)^{2}/2}\langle E'|\psi,t\rangle\langle\psi,t|E\rangle.  
\end{eqnarray}
\end{mathletters}

\noindent In Eq. (5.6c), the pure density matrix of standard quantum theory is multiplied by a gaussian, 
so that $\langle E'|\rho|E\rangle\approx 0$ if $|E'-E|>\hbar/{\cal T}$.  Thus 
one may expect that behavior in standard quantum theory 
which depends upon a coherent superposition of energy states with a spread $> \hbar/{\cal T}$ 
will be significantly altered under energy-driven collapse. 

	Next, consider the behavior of a noninteracting subsystem.  
Denote the subsystem by the subscript 1 and the rest of the universe by the subscript 2.  The initial state is 
$|\psi,0\rangle=|\psi_{1},0\rangle|\psi_{2},0\rangle$ and $H=H_{1}+H_{2}$.  For the ensemble expectation value of 
a quantity $V_{1}$, which depends only upon subsystem 1's variables, Eq. (5.4c) yields:

\begin{mathletters}
\begin{eqnarray}\label{5.7}
\langle V_{1}\rangle(t)=&&{1\over \sqrt{2\pi}}\int_{-\infty}^{\infty}d\eta e^{-\eta^{2}/2}
 \langle\psi_{1},t-\sqrt{\lambda t}\eta |V_{1}|\psi_{1},t-\sqrt{\lambda t}\eta\rangle
 \langle\psi_{2},t-\sqrt{\lambda t}\eta |\psi_{2},t-\sqrt{\lambda t}\eta\rangle\\
 =&&{1\over \sqrt{2\pi}}\int_{-\infty}^{\infty}d\eta e^{-\eta^{2}/2}
 \langle\psi_{1},t-\sqrt{\lambda t}\eta |V_{1}|\psi_{1},t-\sqrt{\lambda t}\eta\rangle.
\end{eqnarray}
\end{mathletters}

	In standard quantum theory, the statevector of two disconnected systems 
describes them as evolving separately. This is not the case with energy-driven collapse: 
the dynamics entangles the systems (through the quadratic dependence on $H$ in the gaussian in Eq. (2.1)).  
However, for the ensemble, the density matrix description has the two systems evolving independently. 
(This interesting result is implicit in the work of references \cite{AdlerHorwitz,Adler}.) 
Since (5.7b) has the same form as (5.4c), the consequences of (5.4c) (smeared pulse behavior, 
unchanging energy spectum, vanishing of sufficiently separated 
off-diagonal elements of the density matrix in the energy representation) 
hold for the subsystem as well. In particular,           
for the ensemble, under the secular energy narrowing to $\Delta E\approx(\lambda t)^{-1/2}$, the energy 
spread allowed to the whole universe is the same energy spread allowed to a noninteracting subsystem!   

\subsection{Individual Behavior} 

	 Now we wish to consider a single universe, characterized by $B(t)$, and 
an experiment set to be performed at time $\approx T$.  For this to make sense, 
the statevector $|\psi,t_{0}\rangle_{B(t_{0})}$ at some time $t_{0}<T$ must be compatible 
with both its function and history. 

	 It is required that 
the apparatus, an isolated subsystem of the universe, 
properly performs the experiment under the standard quantum theory evolution. Thus we hypothesize that 
$|\psi,t_{0}\rangle_{B(t_{0})}=|\psi_{1},t_{0}\rangle|\psi_{2},t_{0}\rangle$, 
where $|\psi_{1},t_{0}\rangle$ describes the subsystem containing the apparatus and $|\psi_{2},t_{0}\rangle$ 
describes the rest of the universe.  That a reasonable initial state of the 
universe, $|\psi,0\rangle$, could have evolved to such a direct product is most unlikely, unless 
energy-driven collapse does indeed take place (so the huge superposition 
corresponding to all the possible universes that $|\psi,0\rangle$  
evolves to under standard quantum theory is collapsed under evolution by $B(t)$ to one of them), 
so this is entailed by the hypothesis.  

	Account must also be taken of the statevector's 
evolution since time 0 (so its energy spectrum 
is appropriately narrrow).  One could consider the evolution of the statevector from $|\psi,t_{0}\rangle_{B(t_{0})}$ 
to $|\psi,t\rangle_{B(t)}$ using Eq. (4.1), but that would not make explicit the limited spectrum of 
$|\psi,t_{0}\rangle_{B(t_{0})}$.  Accordingly, we may employ Eq. (4.4) to display 
the statevector $|\psi,t\rangle_{B(t)}$ in terms of $|\psi,0\rangle$: actually, we shall utilize the 
form (5.3c) which is equivalent to (4.4) (except for the normalization factor).  

	Now, although the integral over $\eta$ in Eq. (5.3c) ranges over $(-\infty, \infty)$ so that 
the statevector $|\psi,t\rangle_{B}$ is formally a superposition of 
statevectors $|\psi,t'\rangle$ for all $t'$, because of the 
weighting factor $\exp-\eta^{2}/4$, in practice the contribution of 
$|\psi,t'\rangle$ for $t'<<t-k{\cal T}$ and $t'>>t+k{\cal T}$ is negligible (where $k$ is chosen so that 
$\exp-k^{2}/4\approx 0$ to the accuracy one wishes). With the above assumption 
of the unentangled nature of the subsystem, the statevector of the 
universe in standard quantum theory may be written as 
$|\psi_{1},t'\rangle|\psi_{2},t'\rangle$ over the interval $T-k{\cal T}\leq t'\leq T+k{\cal T}$.  
It then follows from putting 
the statevector expression Eq. (5.3c) into Eq. (2.3) that  

\begin{eqnarray}\label{5.8}
\langle V_{1}\rangle_{B}(t)
=&&{1\over 4\pi \thinspace_{B}\langle \psi,t|\psi,t\rangle_{B}}
\int d\eta e^{-\eta^{2}/4} \int d\eta' e^{-\eta'^{2}/4}
e^{-iB(t)(\eta-\eta ')/\sqrt{4\lambda t}}\nonumber\\
&&\thinspace\thinspace\thinspace\thinspace\thinspace\thinspace\thinspace\thinspace\thinspace
\cdot \langle\psi_{1},t-\sqrt{\lambda t}\eta ' |V_{1}|\psi_{1},t-\sqrt{\lambda t}\eta\rangle
\langle\psi_{2},t-\sqrt{\lambda t}\eta'|\psi_{2},t-\sqrt{\lambda t}\eta\rangle.
\end{eqnarray}

	The universe is a very big place, with lots going on.  Accordingly, we expect that 
$|\psi_{2},t\rangle$ is orthogonal to $|\psi_{2},t'\rangle$ for $t'$ just slightly different from t 
(we assume that $T$ is large enough so that 
the time scale for the universe to change in standard quantum theory is much shorter than ${\cal T}$).  
That is, it is an excellent aproximation to take 
$\langle\psi_{2},t-\sqrt{\lambda t}\eta'|\psi_{2},t-\sqrt{\lambda t}\eta\rangle\approx c(t)\delta(\eta-\eta ')$.  
Then Eq. (5.8) becomes 

\begin{equation}\label{5.9}
\langle V_{1}\rangle_{B}(t)={1\over \sqrt{2\pi}}\int_{-\infty}^{\infty}d\eta e^{-\eta^{2}/2}
 \langle\psi_{1},t-\sqrt{\lambda t}\eta |V_{1}|\psi_{1},t-\sqrt{\lambda t}\eta\rangle
\end{equation}

\noindent using $\langle 1\rangle_{B}(t)=1$ (which follows from 
 Eq. (2.3)), to select the overall normalization factor. 
 
 	Although this is a single universe, we have arrived at the density matrix description of Eq. (5.9) 
 because the evolution has entangled the subsystem with the rest 
 of the universe which is traced over. When the apparatus 
 is included in the statevector description, if proper collapse occurs, just one of the 
 macroscopically distinguishable outcomes of the experiment 
 should be described by the statevector. However, because one does not 
 know the state of the rest of the universe (which may have a decisive 
 influence on the experiment's outcome), one must use the 
 density matrix description. Then, the density matrix should be diagonal in the basis 
 describing the different outcomes of the experiment, and  
 the probabilities associated with these diagonal states should be interpreted  
 as giving the statistics of these actualized outcomes. If, however, the diagonal states 
 of the density matrix turn out to be inappropriate for the theory 
 (as occurs for the examples in section VI, where each state 
 has more energy spread than is allowed in the universe at time $T$), 
 one is forced to the alternative explanation. It is that 
 proper collapse does not occur: the diagonal density matrix form is there because  
 the statevector describes an apparatus in a superposition 
 of different measurement results entangled with the rest of the universe 
 (which, when the latter is traced over, has the 
 result of cancelling the off-diagonal density matrix elements) as happens in standard quantum theory,

	Yet another way to obtain Eq. (5.9) is to write $\langle V_{1}\rangle_{B}(t)$ in Eq. (5.8) in the form
	 
\begin{eqnarray}\label{5.10}	
\langle V_{1}\rangle_{B}(t)\sim&&\int dE_{2}dE_{1}'dE_{1}|\langle E_{2}|\psi_{2},t\rangle|^{2}
\langle\psi_{1},t|E_{1}'\rangle\langle E_{1}'|V_{1}|E_{1}\rangle\langle E_{1}|\psi_{1},t\rangle\nonumber\\
\thinspace &&\thinspace\thinspace\thinspace\thinspace\thinspace\thinspace\thinspace
\cdot e^{-{1\over 4\lambda t}[B(t)-2\lambda t (E_{2}+E_{1}')]^{2}}
e^{-{1\over 4\lambda t}[B(t)-2\lambda t (E_{2}+E_{1})]^{2}}
\end{eqnarray}

\noindent which is obtained by expressing the statevectors in (5.8) 
in the energy basis, extracting the $\eta$ dependence using 
$\langle E|\psi,t-\sqrt{\lambda t}\eta\rangle=\exp iE\sqrt{\lambda t}\eta\langle E|\psi,t\rangle$ 
and performing the integrals over $\eta$, $\eta '$. 
Since $|\psi_{2},t\rangle$ describes the universe (minus the small subsystem 1), we may take 
$|\langle E_{2}|\psi_{2},t\rangle|^{2}$ in Eq. (5.10) to be approximately constant over the 
ranges of $E_{1}$, $E_{1}'$, $E_{2}$ where the gaussian exponents are small.  Then the integral over $E_{2}$ results in 

\begin{equation}\label{5.11}	
\langle V_{1}\rangle_{B}(t)\sim\int dE_{1}'dE_{1}
\langle\psi_{1},t|E_{1}'\rangle\langle E_{1}'|V_{1}|E_{1}\rangle\langle E_{1}|\psi_{1},t\rangle
e^{-{\lambda t\over 2}(E_{1}'-E_{1})^{2}}.  
\end{equation}

\noindent Note that $B(t)$ has disappeared from Eq. (5.11).  Putting 

\[ e^{-{\lambda t\over 2}(E_{1}'-E_{1})^{2}}={1\over \sqrt{2\pi}}\int d\eta e^{-\eta^{2}/2}
e^{-i\sqrt{\lambda t}\eta(E_{1}'-E_{1})}
\]

\noindent into Eq. (5.11) and choosing the normalization factor so that $\langle 1\rangle_{B}(t)=1$ 
results in (5.9).

	Since (5.9) has the same form as (5.4c), the consequences of (5.4c) (smeared pulse behavior, 
unchanging energy spectum, approximate vanishing of sufficiently separated  
off-diagonal elements of the density matrix in the energy representation) 
hold in this case as well.

 	Lastly, we note that Eq. (5.9) also desribes the situation where  
the initial statevector can be written as $|\psi,0\rangle=|\psi_{1},0\rangle|\psi_{2},0\rangle$ 
where subsystem 1, under standard quantum theory, does not interact the rest of the universe and
describes the apparatus set to turn on at time $T$. It is reassuring to see that,  
under energy-driven collapse, whether the 
apparatus somehow miraculously appears at the start of the universe or is 
constructed at a later time, it has the same description (which, 
of course, is a property of standard quantum theory).  

\section{Two Experiments}\label{VI}
 
	We now analyze two different microscopic phenomena under energy-driven collapse, the precession of 
a spin 1/2 particle in a magnetic field, and excitation of a bound state 
followed by its decay. Each is considered to be an 
isolated system in a single universe, so Eq. (5.9) applies. 
The energy spectrum in 
standard quantum theory in the former case consists of two values, in the latter it is spread over a continuum.  In 
both cases the energy spread (characteristic time) is chosen $>>\hbar/{\cal T}$ ($<<{\cal T}$) so that 
the behavior expected from standard quantum theory is 
appreciably altered.  Under energy-driven collapse, the spin 1/2 particle does not precess,  
while the excitation/decay products do not 
have an exponential time distribution and have a characteristic time ${\cal T}$.    

	If an apparatus were to properly interact with the microscopic 
system, it ought to record this altered behavior.  However, when we 
apply Eq. (5.9) to the combined system+apparatus, we find that the density matrix describes the 
apparatus as recording precession in the first case and recording exponential 
decay in the second case. 

\subsection{Spin Precession} 

	Since application of Eq. (5.9) to an experimental situation has to 
respect the terms under which it was obtained, we need to arrange that 
the experiment commences in the neighborhod of time $T$.  Therefore, in our model, the statevector 
$|\psi_{1},t\rangle$ describes a (one dimensional) ``photon" (spatial coordinate $Q$, momentum $P$) 
which, at time $T$, switches on a ``magnetic field", i.e., gives a spin (described by the Pauli 
matrices ${\bf \sigma}$) the energy $(\epsilon/2)\sigma_{3}$.

	We first discuss this model in standard quantum theory, 
in order to obtain $|\psi_{1},t\rangle$ to use in Eq. (5.9) and 
see what happens under energy-driven collapse.  

The Hamiltonian is

\begin{equation}\label{6.1}
H=P+\Theta (Q)(\epsilon /2)\sigma_{3}.  
\end{equation}

\noindent In Eq. (6.1), for simplicity, we have set the 
energy of the photon to be $P$ rather than $|P|$: then, 
from the Heisenberg equation, $dQ/dt=1$, the photon packet can only move to the right.  
We shall suppose that the part of the apparatus which produces the photon (which we do not model) 
makes its initial state a narrowly localized packet. 
The step function  $\Theta (Q)$ acts like a switch, giving the spin up/down 
states the energy difference $\epsilon$ (chosen large enough so that $\epsilon>>\hbar/{\cal T}$) 
after the photon packet passes the origin at time $\approx T$. 

	The energy eigenstates are 
	
\begin{equation}\label{6.2}
\langle q|E_{\pm}\rangle=e^{ikq}[\Theta (-q)e^{\pm i(\epsilon/2)q}+\Theta (q)]|\pm\rangle
\end{equation}	
	
\noindent with energy eigenvalues $E_{\pm}=k\pm \epsilon/2$, so the 
general solution of Schr\"odinger's equation is	the superposition (writing $s\equiv t-T$) 

\begin{equation}\label{6.3}
\langle q|\psi_{1},t\rangle = \sum_{n=\pm}\int dk e^{i[k+n(\epsilon/2)](q-s)}
[\Theta (-q)+\Theta (q)e^{-in(\epsilon/2)]q}]f_{n}(k)|n\rangle.  
\end{equation}

\noindent By choosing $f_{+}=a(2\pi^{3}/\sigma^{2})^{-1/4}\exp -[k+(\epsilon/2)]^{2}\sigma^{2}$ and 
$f_{-}=b(2\pi^{3}/\sigma^{2})^{-1/4}\exp -[k-(\epsilon/2)]^{2}\sigma^{2}$, 
where $a$ and $b$ are complex constants, $|a|^{2}+|b|^{2}=1$, we obtain a wavefunction 
satisfying the correct initial conditions: 

\begin{equation}\label{6.4}
\langle q|\psi_{1},t\rangle = {1\over (2\pi \sigma^{2})^{1/4}} e^{-(q-s)^{2}/4\sigma^{2}}
\big [\big (a|+\rangle+b|-\rangle\big)\Theta (-q)+
\big (ae^{-i(\epsilon/2)q}|+\rangle+be^{i(\epsilon/2)q}|-\rangle\big)\Theta (q)\big].  
\end{equation}

	According to Eq. (6.4) the spin does not precess for $s<0$, while the photon packet travels toward 
the switch at $q=0$.  If the packet is sufficiently narrow, $\sigma\epsilon<<1$ (which we assume), 
although the $\sim\Theta (q)$ terms in (6.4) are spatially, not temporally, dependent, 
it nonetheless follows from (6.4) that the spin precesses once the photon passes the switch, since then 

\[ 
e^{-(q-s)^{2}/4\sigma^{2}}e^{\pm i(\epsilon/2)q}=
e^{-(q-s)^{2}/4\sigma^{2}}e^{\pm i(\epsilon/2)(q-s)}e^{\pm i(\epsilon/2)s}\approx 
e^{-(q-s)^{2}/4\sigma^{2}}e^{\pm i(\epsilon/2)s}.  
\] 

	Using Eq. (6.4) we calculate (for simplicity, choosing $a$ and $b$ to be real in Eq. (6.5b)
so that the spin is initially in the $x$-$z$ plane, as well as utilizing $\sigma\epsilon<<1$):  	

\begin{mathletters}
\begin{eqnarray}\label{6.5}
\langle \psi_{1},t|\sigma_{1}|\psi_{1},t\rangle =&& (a^{*}b+ab^{*})\Phi(-{s\over \sigma})+
e^{-\epsilon^{2}\sigma^{2}/2}[a^{*}b e^{i\epsilon s}\Phi({s\over \sigma}+i\epsilon\sigma)
+ab^{*} e^{-i\epsilon s}\Phi({s\over \sigma}-i\epsilon\sigma)]\\
\approx && 2ab[\Phi(-{s\over \sigma})+\cos(\epsilon s) \Phi({s\over \sigma})].  
\end{eqnarray}
\end{mathletters}

\noindent $\Phi(x)\equiv (2\pi)^{-1/2}\int_{-\infty}^{x} dy\exp-(y^{2}/2)$ is the so-called normal distribution function,  
sort of a gradual step function, making its transition from 0 to 1 over the range, say, of $|x|\leq 2$ 
(since $\Phi(-2)\approx .02$, $\Phi(2)\approx .98$). 

	Eq. (6.5) shows that the spin does not precess for large negative 
$s/\sigma$, that it does precess for large positive $s/\sigma$, and indicates that the transition 
takes place over, say, $|s|<2\sigma$. For $|s|>2\sigma$, the density matrix corresponding to 
the result (6.5b) is that of pure precession:

\begin{equation}\label{6.6}
Tr_{q}\rho\approx [ae^{-i\epsilon s/2}|+\rangle + be^{i\epsilon s/2}|-\rangle]
[ae^{i\epsilon s/2}\langle+| + be^{-i\epsilon s/2}\langle-|]. 
\end{equation}

\noindent An apparatus which verifies the precession could consist of a clock 
with time resolution $<<{\cal T}$ which triggers a device to ``instantaneously" (i.e., over a period 
better than the clock's resolution), nondestructively and repeatedly measure the spin state at 
those times predicted by (6.6) when it 
comes around to point in 
a particular direction (the device axis). Then, each time it is measured, 
the spin will always be seen parallel to the device axis.  Moreover suppose that,  
if the device sees the spin parallel to its axis for a preset number of spin revolutions, 
the apparatus automatically prints a $\surd$ to indicate that the spin precesses.      

	We now put Eq. (6.5a) into (5.9), to see how the precession fares under energy-driven collapse 
(again, taking $a$ and $b$ to be real and utilizing $\sigma\epsilon<<1$   
as well as $\sigma<<{\cal T}$, in going from (6.7a) to (6.7b)):  	
	
\begin{mathletters}
\begin{eqnarray}\label{6.7}
\langle \sigma_{1}\rangle_{B}(t) =&& (a^{*}b+ab^{*})\Phi(-{s\over \sqrt{\sigma^{2}+{\cal T}^{2}}})+
e^{-\epsilon^{2}(\sigma^{2}+{\cal T}^{2})/2}
[a^{*}b e^{i\epsilon s}\Phi({s+i\epsilon\sigma(\sigma+{\cal T})\over \sqrt{\sigma^{2}+{\cal T}^{2}}})
+ab^{*} e^{-i\epsilon s}\Phi({s-i\epsilon\sigma(\sigma+{\cal T})\over \sqrt{\sigma^{2}+{\cal T}^{2}}}]\\
\approx && 2ab[\Phi(-{s\over {\cal T}})+
e^{-\epsilon^{2}{\cal T}^{2}/2}\cos(\epsilon s) \Phi({s\over {\cal T}})].  
\end{eqnarray}
\end{mathletters}

\noindent Eq. (6.7b) shows that the ``switchover" takes a longer time, i.e., $|s|<2{\cal T}$ instead of 
$|s|<2\sigma$.  This is because the  photon packet is widened: the narrowed 
energy spectrum can no longer sustain a rapid transition. 

	For $|s|>2{\cal T}$, the density matrix corresponding to 
the result (6.7b) may be written as 

\begin{equation}\label{6.8}
Tr_{q}\rho\approx (1-e^{-\epsilon^{2}{\cal T}^{2}/2})[a^{2}|+\rangle\langle+|+b^{2}|-\rangle\langle-|] 
+e^{-\epsilon^{2}{\cal T}^{2}/2}[ae^{-i\epsilon s/2}|+\rangle + be^{i\epsilon s/2}|-\rangle]
[ae^{i\epsilon s/2}\langle+| + be^{-i\epsilon s/2}\langle-|],
\end{equation}

\noindent a mixture of spin up, spin down and spin precessing. 
However, the precession part of the mixture 
becomes negligibly small if we take the energy separation $\epsilon>>\hbar/{\cal T}$.
Therefore, in this case, if the apparatus we have discussed should encounter the 
spin described by (6.8), the repeated measurements should produce 
the result that the spin is sometimes parallel and sometimes antiparallel 
to the device axis, and the apparatus will not print a $\surd$.  

	Now, lets consider what happens according to energy-driven collapse 
when the apparatus is included in the statevector $|\psi_{1},t\rangle$ and 
Eq. (5.9) is applied. Eq. (5.9) gives the density matrix at time $t$ as a 
superposition of pure states, 
each of which describes the standard quantum evolution at a certain instant of time, 
in a $\approx{\cal T}$ neighborhood of $t$. Since, under the standard quantum evolution, there is precession  
which is measured by the apparatus, for $t-T$ sufficiently large 
every apparatus in the superposition has printed out a $\surd$.   
Thus the density matrix at a large enough time $t$ eventually describes that 
a $\surd$ is printed out with certainty.  

	So, we have a paradox.  Analysis of the microscopic behavior indicates that there is no precession. 
Analysis of the microscopic system in interaction with the apparatus indicates that 
there is precession.  In both cases, we have correctly applied Eq. (5.9) to the relevant 
situation.

	To see what has gone wrong, consider the statevector $|\psi_{1},t-\eta {\cal T}\rangle$ 
which contributes to the density matrix integral in (5.9) (for $\eta$ of the order of a small integer).  
This statevector is 
the direct product of the apparatus state (which has recorded a certain number of spin detections 
parallel to its axis) including the clock which reads time $t-\eta {\cal T}$, and the narrow photon  
packet (which had earlier triggered the precession) at that time, 
and the spin state at that time.  Since the 
the clock is so precise, and the photon packet is so narrow, these states 
are orthogonal for fairly close values of $\eta$, each state describing a different clock time 
and a different photon packet location (as well as a different spin orientation which, 
however, is not responsible for the orthogonality). As a result, the 
integral in (5.9) gives a density matrix with the various distinguishable 
$|\psi_{1},t-\eta {\cal T}\rangle\langle \psi_{1},t-\eta {\cal T}|$ 
along the diagonal, to a good approximation. Then, as discussed following Eq. (5.9), 
if collapse occurs properly, a diagonal state should be interpreted as 
a possible outcome of the experiment in the single universe.  However, such 
a state contains the localized photon packet of energy spread $>>{\cal T}^{-1}$ (not to 
mention the clock for which a similar statement obtains), more energy 
than is allowed in the whole universe. Therefore, the density matrix's diagonal states 
cannot be interpreted in this way.  Instead, as discussed, they arise from a statevector 
which describes macroscopically different apparatus states entangled with the rest of the universe.
The apparatus state does not properly correlate to the microscopic state 
because the statevector does not collapse properly.  

\subsection{Excitation of a Bound State and Its Decay}

	Next, we consider a (one-dimensional) ``photon" which at time $\approx T$ excites a bound state 
(located at $x_{0}$) which subsequently decays.  The Hamiltonian is a modification of 
a well known model of a two-state atom\cite{Eberly}:
	 
\begin{equation}\label{6.9}
H=\epsilon b^{\dagger}b+\int_{-\infty}^{\infty} dk k a_{k}^{\dagger} a_{k} +
g\int_{-\infty}^{\infty} dk[a_{k}e^{ikx_{0}}b^{\dagger}+a_{k}^{\dagger}e^{-ikx_{0}}b].   
\end{equation}

\noindent Here $b^{\dagger}$ creates the excited state of energy $\epsilon$ ($[b,b^{\dagger}]=1$) 
and $a_{k}^{\dagger}$ creates a photon of momentum $k$ ($[a_{k}, a_{k'}^{\dagger}]=\delta (k-k')$). 
The coupling constant $g=(\Gamma/2\pi)^{1/2}$, where $\Gamma$ turns out to be the 
bound state lifetime. As in the previous subsection, 
for simplicity, we choose the photon energy to be $k$ rather than $|k|$, 
so that the photon only moves to the right but, also,   
the consequent unbounded energy spectrum allows the 
decay to be precisely exponential\cite{Larry}. 

	First, the analysis in standard quantum theory.  The statevector has the form 

\begin{equation}\label{6.10}
|\psi,t\rangle = \beta (t) b^{\dagger}|0>+\int_{-\infty}^{\infty} dk\alpha_{k}(t) a_{k}^{\dagger}|0> 
\end{equation}

\noindent where $|0\rangle$ is the no-photon state and the ground state of the bound state.  The 
Schr\"odinger equation implies

\begin{equation}\label{6.11}
id \alpha_{k}/dt =g\beta e^{-ikx_{0}}+k\alpha_{k}, \thinspace\thinspace\thinspace
id \beta/dt =\epsilon \beta + g\int_{-\infty}^{\infty} dk\alpha_{k}e^{ikx_{0}}.  
\end{equation}

	For insight, it is worth checking out the solution of Eq. (6.11) for decay without excitation,  
(even though its initial conditions $\beta(T)=1$, $\alpha_{k}(T)=0$ 
are inappropriate for use in Eq. (5.9) if $\Gamma ^{-1}<{\cal T}$, since then 
the initial state will have more energy than is allowed to the universe at time $T$).  
  Setting $s\equiv t-T$, the result for $s>0$ is
 
\begin{mathletters}\label{6.12}
\begin{eqnarray}
\alpha_{k}(t)=&& ge^{-ikx_{0}}{-e^{-((\Gamma /2)+i\epsilon) s}+e^{-iks}\over k-\epsilon +i(\Gamma /2)}\\
\beta(t)=&&e^{-((\Gamma /2)+i\epsilon) s}. 
\end{eqnarray}
\end{mathletters}

\noindent From Eqs. (6.12), the expectation value of the photon number density 
and the expectation value of the particle being in the excited state are 

\begin{mathletters}\label{6.13}
\begin{eqnarray}
\langle\psi,t|a_{k}^{\dagger}a_{k}|\psi,t\rangle =&& {2\pi\Gamma\over(k-\epsilon)^{2}+
(\Gamma /2)^{2}}[1+e^{-\Gamma s}-2e^{-(\Gamma/2) s}\cos(k-\epsilon)s]\\
\langle\psi,t|b^{\dagger}b|\psi,t\rangle =&&e^{-\Gamma s}. 
\end{eqnarray}
\end{mathletters}

\noindent According to Eq. (6.13b), the decay is exponential.  According to Eq. (6.13a), 
the photon distribution is Lorentzian for large $s$.  From Eqs. (6.10), (6.12a), the photon wavefunction 
in the position representation $|x\rangle =(2\pi)^{-1/2}\int dk \exp(-ikx)a_{k}^{\dagger}|0\rangle$ is found:

\begin{equation}\label{6.14}
\langle x|\psi,t\rangle=i\Gamma^{1/2}e^{-i\epsilon [s-(x-x_{0})])}e^{-(\Gamma /2)[s-(x-x_{0})]}
[\Theta(x-x_{0}-s)-\Theta(x-x_{0})].
\end{equation} 

\noindent Thus the photon packet emerging from the decay has 
$|\langle x|\psi,t\rangle|^{2}$ exponentially rising from the value $\Gamma\exp-\Gamma s$ 
at $x=x_{0}$ to $\Gamma$ at $x=x_{0}+s$ and vanishing elsewhere. (In this, and all 
the rest of the examples discussed, it can readily be checked that probability is conserved.)   
 
	In the case of interest, excitation of the bound state followed by decay, the initial conditions are 
$\beta(t)\approx 0$ for $s<0$, and an initial photon wavepacket which reaches  
$x_{0}$ at $s\approx 0$. The solution of Eqs. (6.11) is then 

 \begin{mathletters}\label{6.15}
\begin{eqnarray}
\langle x|\psi,t\rangle=&&f[s-(x-x_{0})]-i\Gamma^{1/2}\Theta(x-x_{0})\beta[s-(x-x_{0})]\\
\beta(s)=&&-i(\Gamma)^{1/2}e^{-[(\Gamma/2)+i\epsilon]s}\int_{-\infty}^{s} ds'f(s')e^{[(\Gamma/2)+i\epsilon]s'}
\end{eqnarray}
\end{mathletters}

\noindent where $f(z)$ describes the incident wavepacket.
  
	In what follows, we shall take the width of the incident wavepacket, $\sigma$, to be much less than 
any time in the model, so $\sigma\epsilon<<1$, $\sigma\Gamma<<1$ and $\sigma{\cal T}^{-1}<<1$ 
can all be neglected compared to 1.  In that case, we can also take $f(s)$ to be approximately 
 the ``square root of 
a delta function," $f(s)\approx \sigma^{1/2}\delta (s)$ (e.g., the square root of 
a normalized narrow gaussian is a narrow unnormalized gaussian).   Then, the probabilities given 
by Eqs. (6.15) may be written as 

\begin{mathletters}\label{6.16}
\begin{eqnarray}
|\langle x|\psi,t\rangle|^{2}=&&|f[s-(x-x_{0})]|^{2}-\Gamma\sigma \Theta (x-x_{0}) \delta [s-(x-x_{0})] 
+\Gamma^{2}\sigma \Theta (x-x_{0}) \Theta [s-(x-x_{0})] e^{-\Gamma [s-(x-x_{0})]}\\
|\beta(s)|^{2}\approx&&\Gamma\sigma \Theta (s) e^{-\Gamma s}.   
\end{eqnarray}
\end{mathletters}

	According to Eq. (6.16b), $|\beta(s)|^{2}=0$ for negative $s$, jumps to $\Gamma\sigma$ at 
$s=0$, and thereafter exponentially decays with lifetime 
$\Gamma$.  According to Eq. (6.16a), the squared photon wavefunction consists of 
the initial wavepacket, an interference term between it and the 
leading edge of the decaying packet and the decay product whose square vanishes 
for $x<x_{0}$,      
jumps at $x=x_{0}$ (when $s>0$) to $\Gamma^{2}\sigma\exp-\Gamma s$, exponentially 
rises as $x$ increases to $\Gamma^{2}\sigma$ at $x=x_{0}+s$ and vanishes for $x>x_{0}+s$.

	Now lets see what happens under energy-driven collapse.  Using 
Eqs. (5.9) and (6.15), we calculate the expectation value of the photon position probability distribution 
and the expectation value of the occupation number of the excited state: 

 \begin{mathletters}\label{6.17}
\begin{eqnarray}
	\langle |x\rangle\langle x|\rangle_{B}(t)=&&(2\pi)^{-1/2}
\int_{-\infty}^{\infty}d\eta e^{-\eta^{2}/2}|\langle x|\psi,T-{\cal T}\eta\rangle|^{2}\nonumber\\
=&&(2\pi {\cal T}^{2})^{-1/2}e^{-{1\over2{\cal T}^{2}}[s-(x-x_{0})]^{2}}[1-
\Gamma\sigma \Theta (x-x_{0})]\nonumber\\
&&\thinspace\thinspace\thinspace\thinspace\thinspace\thinspace\thinspace\thinspace\thinspace\thinspace
+\Gamma^{2}\sigma \Theta (x-x_{0})\Phi[(s-(x-x_{0})/{\cal T})-\Gamma{\cal T}]
e^{-\Gamma [s-(x-x_{0})]}e^{(1/2)(\Gamma{\cal T})^{2}}\\
\langle b^{\dagger}b\rangle_{B}(t)=&&(2\pi)^{-1/2}
\int_{-\infty}^{\infty}d\eta e^{-\eta^{2}/2}|\beta(t-{\cal T}\eta)|^{2}\nonumber\\
=&&\Gamma\sigma e^{-\Gamma s}e^{(1/2)(\Gamma{\cal T})^{2}}\Phi[(s/{\cal T})-\Gamma{\cal T}].
\end{eqnarray}
\end{mathletters}

	First consider the small ${\cal T}$ case, i.e., ${\cal T}\Gamma<<1$.  From Eq. (6.17b),  
the bound state excitation onset, governed by $\approx\Phi[s/{\cal T}]$, takes place over ${\cal T}$, 
because the incident photon wavepacket width is broadened from $\sigma$ to 
${\cal T}$. The decay is exponential 
with time constant $\Gamma^{-1}$.  From Eq. (6.17a), 
the interference term at the leading edge of the outgoing packets has likewise been broadened to 
a gaussian of width $\Gamma^{-1}$. 
In other words, the outgoing packet's behavior is similar to that of 
standard quantum theory's (6.16b), except that its 
onset takes place over ${\cal T}$.

	Of greatest interest is the large ${\cal T}$ case. Utilizing the large $x$ behavior 
$\Phi[-x]\rightarrow (2\pi)^{-1/2}x^{-1}\exp-(x^{2}/2)$, and ${\cal T}\Gamma>>1$, Eqs. (6.17) becomes 

 \begin{mathletters}\label{6.18}
\begin{eqnarray}
	\langle |x\rangle\langle |x\rangle_{B}(t) 
\approx &&(2\pi {\cal T}^{2})^{-1/2}e^{-{1\over2{\cal T}^{2}}[s-(x-x_{0})]^{2}}
\bigg\{ [1-\Gamma\sigma \Theta (x-x_{0})]
+\Gamma\sigma \Theta (x-x_{0})\bigg[1+{s-(x-x_{0})\over \Gamma{\cal T}^{2}}\bigg]\bigg\}\\
\langle b^{\dagger}b\rangle_{B}(t)\approx &&\sigma (2\pi {\cal T}^{2})^{-1/2}e^{-{1\over2{\cal T}^{2}}s^{2}}. 
\end{eqnarray}
\end{mathletters} 

		From Eq. (6.18b), the decay of the bound state is no longer exponential. It is gaussian with 
characteristic time ${\cal T}$.  
This is because the incident wavepacket is gaussian, not because the decay is gaussian:  
Eqs. (6.18) behave just like the standard quantum theory description Eqs. (6.15) with a broad 
gaussian incident wavepacket, of width ${\cal T}$, and a relatively rapid decay ($\Gamma^{-1}<<{\cal T}$).  
Although the decay is faster than  ${\cal T}$, it does not appear in any equation since the decay is 
masked by the ${\cal T}$ behavior of the incident wavepacket. Thus there is no
violation of the stricture against ``fast" pulse behavior.   

	From Eq. (6.18a), the interference term (second term in the small square bracket) 
and most of the outgoing packet (first term in 
the large square bracket) cancel, so that what remains has the shape of the incident packet  plus 
a little ``blip" (second term in 
the large square bracket), which is a little larger at the leading edge and a little smaller at the trailing edge. 

	While, in this model, the incident photon 
and the photon decay product cannot be separated (and $\Gamma$ 
completely disappears from Eqs. (6.18)), if one wishes one can alter the model by 
adding to the Hamiltonian (6.9) another coupling term:  

\[ g'\int_{-\infty}^{\infty} dk [c_{k}e^{ikx_{0}}b^{\dagger}+c_{k}^{\dagger}e^{-ikx_{0}}b]\]

\noindent so that the excited state can decay to a $c$-particle as well as an $a$-particle.  
Then, the decay product can be separated from 
the incident particle by making the decay go essentially only to the $c$-particle: 
choose $g\sim \Gamma^{1/2}$ small enough ($\Gamma^{-1}>>{\cal T}$),  
and $g'\sim \Gamma'^{1/2}$ large enough ($\Gamma'^{-1}<<{\cal T}$), so the state becomes excited 
by the incident $a$-particle, but decays to the $c$-particle. 
Then, under energy-driven collapse, the large square bracket term in (6.18a) will describe the $c$-particle decay product.

	Suppose an accelerator produces a localized burst of many such photons and they 
hit many such bound states, spread out in a thin film of thickness $\sigma$ so that, in 
the standard quantum theory description, the bound states are excited essentially simultaneously.  Suppose 
a detector, with an accurate clock (of time resolution much better than ${\cal T}$) 
can measure the time distribution of arrival of the resulting 
outgoing particles. For use shortly, we shall assume the measurement is reasonably 
nondisturbing, so a detected particle is allowed to proceed beyond the detector.
Moreover, suppose the apparatus is designed to print out, at its leisure, the time spread of the 
decay products and whether the decay shape is exponential ($\cal YES$) or not ($\cal NO$). 
Then, a properly operating detector encountering 
many photons, each described by Eqs. (6.18), should print out ${\cal T}$ and ${\cal NO}$.  

	Now, suppose the detector is included in the statevector. In the standard 
quantum description, the detector sees the results described in Eqs. (6.16) and 
will, at its leisure, print out $\Gamma$ and $\cal YES$. Then, to see what happens under energy-driven collapse,  
we apply Eq. (5.9). The density matrix is a superposition of pure states 
$|\psi, t-{\cal T}\eta\rangle\langle \psi, t-{\cal T}\eta |$, each of which 
satisfies the standard quantum description.  Therefore, for $t$ large enough, 
the density matrix describes with certainty that the apparatus prints out $\Gamma$ and $\cal YES$.
   
	Thus, as in the previous subsection we have a paradox.  There is a conflict between the 
result for the microscopic system alone, and the result for 
the apparatus interacting with the microscopic system, both correctly calculated 
using Eq. (5.9). 

	To see what has gone wrong, consider the nature of the density matrix at some time 
$t$ during the measurement, calculated according to Eq. (5.9).   Standard quantum theory does not produce collapse, so the 
statevector in standard quantum theory at time  $t-{\cal T}\eta$ is   
a superposition of states, each describing an apparatus which 
has recorded a particular photon detection sequence in a direct product with 
the photons which have been nondestructively detected.  Although 
the conclusion from each sequence is the same and leads to the same summarizing printout, since all sequences are different 
these are macroscopically distinguishable states. 
Under the time-smearing construction of (5.9), these states appear superposed in the resulting density matrix. 
If the density matrix is not diagonal in these states then, obviously, the superposition of 
such states which occurs in standard quantum theory has carried over to 
the energy-driven collapse theory, and collapse is not working.  Suppose, then, that 
the density matrix is diagonal in these states.    
These are states for which the photons, resolved to better than $\Gamma^{-1}$, 
have more energy than is allowed in the universe (not to mention the apparatus's clock for which 
a similar statement obtains). Thus again, as discussed following Eq. (5.9) and in the last subsection, 
the density matrix could not be the result of proper collapse.  

\subsection{Numerical Values}

	We deliberately have not 
put numbers into these discussions of experiments. The point we are making, that 
proper collapse does not occur when the apparatus is included in the statevector, 
is a point of principle, not a conflict with experiment. However, it is worth considering a few examples of 
altered phenomena which are predicted when the apparatus is not included in the statevector.
	
	Hughston has made the interesting choice $\lambda=(G/\hbar^{3}c^{5})^{1/2}$=  
(Planck time)/$\hbar^{2}$ for which, with $T=13.7$billion years, one obtains 
${\cal T}\equiv (\lambda T)^{1/2}\approx 1.5\cdot 10^{-13}$sec. However, because timers, counters, oscilloscopes, printers etc. 
don't operate at $10^{-13}$sec, it is hard to come up with an experiment actually performed or even presently performable,
for which the slight smearing over $10^{-13}$sec has a practical detectable effect. 

	For example, there has been much work with pulses of 
$\approx$1 to 100 femtoseconds (1fs=$10^{-15}$sec) which can be obtained from a mode-locked Ti:sappphire laser.  
In one such experiment, the cross-correlation 
of the intensities of two such pulses is measured\cite{Meshulach}. A 27fs pulse (centered at 800nm) is split into two pulses. 
One pulse is reshaped, the other delayed, and the two are recombined in a frequency-doubling crystal.  The output intensity,
filtered at twice the input frequency, is proportional to the intensity cross-correlation of the two pulses 
at the delay time, and is measured by a photomultiplier tube. In one case, 
graphs of intensity cross-correlation versus delay time show <54fs structure.  Nonetheless, the description 
under energy-driven collapse which causes a 
150fs ``smearing" gives no different result than standard quantum theory because  
both pulses are similarly affected.  The averaging of 
$|\psi_{1},t-{\cal T}\eta\rangle\langle \psi_{1},t-{\cal T}\eta|$ in (5.9) 
just means that the signal proportional to the cross-correlation 
coming out of the frequency-doubling crystal will be spread over $\approx$150fs instead of $<54$fs.  
The photomultiplier measures its input intensity independently of the input pulse width, 
so the measurement result will be the same.  Other experiments, 
such as those involving time-domain terahertz spectroscopy, are similarly configured, 
and thus similarly unaffected.  

	A direct measurement of pulse width, showing that a pulse less than ${\cal T}$ 
in width {\it can} be observed, {\it would} certainly be adequate to produce a discrepancy between the 
energy-driven collapse prediction and experiment. 
The fastest commercial oscilloscope of which I am aware, the Tektronix TDS 6604, 
operates at 6 GHz, with a 20Gigasample/sec rate, so if ${\cal T}\approx 10^{-10}$sec, 
a discrepancy would be observed.

	A similar situation prevails with regard to decay experiments 
such as discussed in Section VIB.  For example, the  $\pi^{0}$'s lifetime of $\approx .1fs$, 
time dilated to $\approx 5fs$ (the pions had energy 7.1GeV), has been 
measured by what may be thought of as a time-of-flight experiment\cite{von Dardel}. 
18GeV protons bombarded platinum foils of various thicknesses, uniformly creating pions in a foil.  
If a moving pion has a short enough lifetime so that it decays to 2 gammas while it is still in the foil, 
there is a certain probability (largest when the gammas still have a lot of foil to travel through) 
that a $\gamma$ will create an electron-positron pair by colliding with an atom in the foil.  The positrons are detected, and  
the number of positrons produced in foils of various thicknesses can be related to 
the pion lifetime. Under energy-driven collapse, the observed results would be no different, since it 
is effectively distances that are measured.   
All energy-driven collapse requires is that the $\pi^{0}$ and decay product wavefunctions 
rise and fall over time $\geq {\cal T}$: there is no effect on the size of the {\it distance} 
travelled before decay or pair creation.   
 
    However, with regard to the spin precession discussion in Section VIA, it is 
presently just possible to obtain a magnetic field strong enough to 
observe an experimental discrepancy.  An electron spin in a magnetic field precesses at 
2.8MHz/Gauss.  The largest pulsed magnetic fields at present, produced by machines in high magnetic field laboratories 
around the world, are $\approx$70T, corresponding to a frequency of $\approx 2\cdot 10^{12}$Hz.  But, a magnetic field 
of 850T, corresponding to a precession frequency of $\approx 2\cdot 10^{13}$Hz has been produced in a 
one-shot ``self-destructive" magnet.  Suppose a slug of matter were to be 
placed between the poles of such a magnet during its few millisec of operation, and the far infrared 
magnetic dipole radiation expected to be produced by the precessing electrons were to be observed.  The diminished 
radiation predicted by energy-driven collapse due to the diminished precession, 
compared to standard quantum theory's prediction of the precession, could be tested.

\acknowledgments

 I would especially like to thank Steve Adler for his hospitality 
 at the Institute for Advanced Study at Princeton 
 where this work was conceived, and for his many helpful comments on this paper.  I would also like to thank 
 Todd Brun, Brian Collett, Larry Horwitz, Lane Hughston, Gordon Jones, Jim Ring, Ann Silversmith and Don Stewart 
 for useful remarks.

\end{document}